# A Contextual Risk Model for the Ellsberg Paradox

Diederik Aerts and Sandro Sozzo
Center Leo Apostel for Interdisciplinary Studies
Brussels Free University, Pleinlaan 2, 1050 Brussels
E-Mails: diraerts@vub.ac.be, ssozzo@vub.ac.be

**Abstract**

The *Allais* and *Ellsberg paradoxes* show that the *expected utility hypothesis* and Savage's *Sure-Thing Principle* are violated in real life decisions. The popular explanation in terms of *ambiguity aversion* is not completely accepted. On the other hand, we have recently introduced a notion of *contextual risk* to mathematically capture what is known as *ambiguity* in the economics literature. Situations in which contextual risk occurs cannot be modeled by Kolmogorovian classical probabilistic structures, but a non-Kolmogorovian framework with a quantum-like structure is needed. We prove in this paper that the contextual risk approach can be applied to the Ellsberg paradox, and elaborate a sphere model within our hidden measurement formalism which reveals that it is the overall conceptual landscape that is responsible of the disagreement between actual human decisions and the predictions of expected utility theory, which generates the paradox. This result points to the presence of a *quantum conceptual layer* in human thought which is superposed to the usually assumed *classical logical layer*.

## 1. The Sure-Thing Principle and the Ellsberg Paradox

The expected utility hypothesis requires that in uncertain circumstances individuals choose in such a way that they maximize the expected value of 'satisfaction' or 'utility'. This hypothesis is the predominant model of choice under uncertainty in economics, and is founded on the *von Neumann-Morgenstern utility theory* (von Neumann and Morgenstern 1944). These authors provided a set of 'reasonable' axioms under which the expected utility hypothesis holds. One of the proposed axioms is the *independence axiom* which is an expression of Savage's *Sure-Thing Principle* (Savage 1954). Examples exist in the literature, which show an inconsistency with the predictions of the expected utility hypothesis, namely a violation of the Sure-Thing Principle. These deviations, often called paradoxes, were firstly revealed by Maurice Allais (Allais 1953) and Daniel Ellsberg (Ellsberg 1961). The Allais and Ellsberg paradoxes at first sight indicate the existence of an *ambiguity aversion*, that is, individuals prefer 'sure' over 'uncertain' choices. Several attempts have been propounded to solve the drawbacks raised by these paradoxes but none of the existing arguments is universally accepted.

Let us analyze these above problems in more details. Savage introduced the *Sure-Thing Principle* (Savage 1954) inspired by the following story.

*A businessman contemplates buying a certain piece of property. He considers the outcome of the next presidential election relevant. So, to clarify the matter to himself, he asks whether he would buy if he knew that the Democratic candidate were going to win, and decides that he would. Similarly, he considers whether he would buy if he knew that the Republican candidate were going to win, and again finds that he would. Seeing that he would buy in either event, he decides that he should buy, even though he does not know which event obtains, or will obtain, as we would ordinarily say.*

The Sure-Thing Principle is equivalent to the assumption that, if persons are indifferent in choosing between simple lotteries $L_1$ and $L_2$, they will also be indifferent in choosing between $L_1$ mixed with an arbitrary simple lottery $L_3$ with probability p and $L_2$ mixed with $L_3$ with the same probability p (*independence axiom*).

Let us now consider the situation proposed by Daniel Ellsberg (Ellsberg 1961) to point out an inconsistency with the predictions of the expected utility hypothesis and a violation of the Sure-Thing Principle. Imagine an urn known to contain 30 red balls and 60 balls that are either black or yellow, the latter in unknown proportion. One ball is to be drawn at random from the urn. To 'bet on red' means that you will receive a prize a (say, 10 euros) if you draw a red ball ('if red occurs') and a smaller amount b (say, 0 euros) if you do not. If test subjects are given the following four options: (I) 'a bet on red', (II) 'a bet on black', (III) 'a bet on red or yellow', (IV) 'a bet on black or yellow', and are then presented with the choice between bet I and bet II, and the choice between bet III and bet IV, it appears that a very frequent pattern of response is that bet I is preferred to bet II, and bet IV is preferred to bet III. This violates the Sure-Thing Principle, which requires the ordering of I to II to be preserved in III and IV (since these two pairs differ only in the pay-off when a yellow ball is drawn, which is constant for each pair). The first pattern, for example, implies that test subjects bet on red rather than on black; and also that they will bet against red rather than against black.

The contradiction above suggests that preferences of real life subjects are inconsistent with the Sure-Thing Principle. A possible explanation of this difficulty could be that people make a mistake in their choice and that the paradox is caused by an error of reasoning. We have recently studied these paradoxes, together with the existing attempts to solve them, and we instead argue that subjects make their decisions violating the Sure-Thing Principle, but not because they make an error of reasoning, but rather because they follow a different type of reasoning. This reasoning is not only guided by logic but also by conceptual thinking which is structurally related to quantum mechanics (Aerts and D'Hooghe 2009). In particular, we have performed in (Aerts, D'Hooghe and Sozzo 2011) a test of the Ellsberg paradox on a sample of real subjects. We have also asked them to explain the motivations of their choices. As a consequence of their answers, we have identified some conceptual landscapes that act as decision contexts surrounding the decision situation and influencing the subjects' choices in the Ellsberg paradox situation. We only report these conceptual landscapes in the following and refer to (Aerts, D'Hooghe and Sozzo 2011) for a complete analysis of them.

(i) *Physical landscape*: 'an urn is filled with 30 balls that are red, and with 60 balls chosen at random from a collection of black and a collection of yellow balls'.

(ii) *First choice pessimistic landscape*: 'there might well be substantially fewer black balls than yellow balls in the urn, and so also substantially fewer black balls than red balls'.

(iii) *First choice optimistic landscape*: 'there might well be substantially more black balls than yellow balls in the urn, and so also substantially more black balls than red balls'.

(iv) *Second choice pessimistic landscape*: 'there might well be substantially fewer yellow balls than black balls, and so substantially fewer red plus yellow balls than black plus yellow balls, of which there are a fixed number, namely 60'.

(v) *Second choice optimistic landscape*: 'there might well be substantially more yellow balls than black balls, and so substantially more red plus yellow balls than black plus yellow balls, of which there are a fixed number, namely 60'.

(vi) *Suspicion landscape*: 'who knows how well the urns has been prepared, because after all, to put in 30 red balls is straightforward enough, but to pick 60 black and yellow balls is quite another thing; who knows whether this is a fair selection or somehow a biased operation, there may even have been some kind of trickery involved'.

(vii) *Don't Bother Me With Complications Landscape*: 'if things become too complicated I'll bet on the simple situation, the one I understand well'.

The results obtained in the experimental analysis we have carried out have allowed us to suggest that it is the combined effect of the above (and, possibly, other) landscapes that is responsible, together with ambiguity aversion, of the experimental results collected since Ellsberg, hence of the deviations from classically expected behavior. The main consequence of the presence of these contextual effects is that a quantum or, better, quantum-like, formalism is needed to model the Ellsberg situation at a statistical level, as we explained in detail in (Aerts, D'Hooghe and Sozzo 2011). This insight will be extensively strengthened and deepened in Secs. 3 and 4 in connection with our hidden measurement formalism and the suggested quantum conceptual layer. But we first introduce the notion of contextual risk, which will be done in the following section.

## 2. Contextual risk within the hidden measurement formalism

The introduction of the notion of contextual risk is presented in great detail in Aerts and Sozzo (2011). We resume here the essentials of it that are needed for attaining our results in Secs. 3 and 4.

Frank Knight introduced an interesting distinction between different kinds of uncertainty (Knight 1921), and Daniel Ellsberg inquired into the conceptual differences between them (Ellsberg 1961). More explicitly, Ellsberg put forward the notion of *ambiguity* as an uncertainty that does not admit a defined probability measure modeling it, as opposed to *risk*, where such a probability measure instead exists. The difference between ambiguity and risk can be grasped at once by considering the situation introduced by Ellsberg himself. Indeed, in this case, 'betting on red' is associated with a defined probability measure, i.e. a probability of 1/3 to win the bet, and a probability of 2/3 to lose it. For 'betting on black', however, there is no definite probability measure. Indeed, since we only know that the sum of the black and the yellow balls is 60, the number of black balls is not known. In absence of any additional information, 'betting on black' is a situation of ambiguity.

Ellsberg implicitly was considering classical, or Kolmogorovian, probability (Kolmogorov 1933) in his definition of risk and ambiguity. The research on the foundations of quantum mechanics has meanwhile shown that classical probability is not the most general conceivable probabilistic framework, since it cannot model situations where context plays a crucial role. It is worth to be more explicit on this point and devote some words to it. In classical physics one can construct models which include indeterminism, e.g., statistical mechanical models. But, this indeterminism only describes the subjective *lack of knowledge* about the pure state in which the physical entity has been prepared. Thus, a notion of *statistical*, or *mixed*, state is introduced. The ensuing probability model satisfies the axioms of Kolmogorov (*Kolmogorovian probability*). The situation is different in quantum mechanics where the probability model involved is non-Kolmogorovian (Accardi 1982, Accardi and Fedullo 1982, Pitowsky 1989). One of the authors has proved that the non-Kolmogorovian nature of quantum probability is due to a lack of knowledge about how context (in the case of quantum mechanics, measurement context) interacts with the entity that is considered, i.e. it is due to the presence of *fluctuations* in the interaction between context and entity. This result has been successively deepened and a *hidden measurement formalism* has been worked out in which it has been shown that, whenever the effects of context on a (not necessarily physical) entity can be neither neglected nor predicted, then the probabilistic framework describing this situation is necessarily non-Kolmogorovian and admits either a pure (Hilbert space) quantum or a quantum-like representation (Aerts 1986, 1993, 1995, 1998, 2002). Cases in which context plays a fundamental role frequently occur in both everyday life and economics. For this reason, we have introduced in a recent paper *contextual risk* to model the context dependent situations that are described in the economic literature in terms of ambiguity (Aerts and Sozzo 2011). The main and pragmatically relevant difference is that the notion of contextual risk can be recovered within our general hidden measurement formalism, hence it can be endowed with a non-Kolmogorovian quantum-like probability structure.

The way in which the hidden measurement formalism can be applied to contextual risk will be evident in the next section where an explicit hidden measurement model will be constructed for the Ellsberg paradox.

## 3. A sphere model for the Ellsberg paradox

In this section we intend to elaborate a macroscopic physical model that reproduces the statistical features of the decision system considered by Ellsberg. Our sphere model is a realization, but also an extension, of the sphere model for contextual risk worked out in (Aerts and Sozzo 2011). We will see that the sphere model presented here has a non-Kolmogorovian quantum-like structure, which provides a theoretical support to maintain that also the Ellsberg system should exhibit the same features.

The sphere model consists of a physical entity $S$ that is a material point particle $P$ moving on the surface of a sphere, denoted by *surf*, with center $O$ and radius 1. The unit vector $v$ where the particle is located on *surf* represents the pure state $p_v$ of the entity $S$ (see Fig. 1a). For each point $u \in surf$, we introduce the following measurement $e_u$. We consider the diametrically opposite point $-u$ and install a piece of elastic of 2 units of length such that it is fixed with one of its endpoints in $u$ and the other endpoint in $-u$. Once the elastic is installed, the material particle $P$ falls from its original place $v$ orthogonally onto the elastic, and sticks on it (Fig. 1b). Then the elastic breaks somewhere and the particle $P$, attached to one of the two pieces of the elastic (Fig. 1c), moves to one of the two endpoints $u$ or $-u$ (Fig. 1d). Depending on whether the particle $P$ arrives at $u$ (as in Fig. 1) or at $-u$, we attribute the outcome $o_1^u$ or $o_2^u$ to $e_u$. The elastic installed between $u$ and $-u$ plays the role of a (measurement) context for the entity $S$.

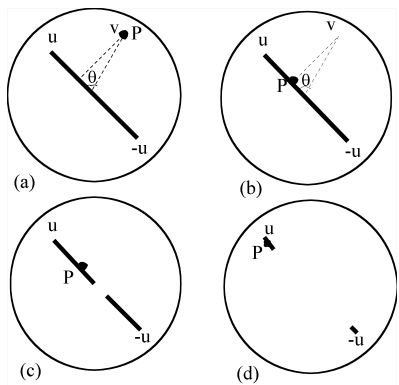

Fig. 1: A representation of the sphere model

Let us now consider elastics that break in different ways depending on their physical construction or on other environmental happenings. We can describe such a general situation by a classical probability distribution

$$\rho : [-u, u] \to [0, +\infty[ \qquad (1)$$

such that

$$\int_\Omega \rho(x) dx \qquad (2)$$

is the probability that the elastic breaks in the region $\Omega \subset [-u, u]$. We also have:

$$\int_{[-u,u]} \rho(x) dx = 1, \qquad (3)$$

which expresses the fact that the elastic always breaks during a measurement. A measurement $e_u$ characterized by $\rho$ will be called a $\rho$-*measurement* and denoted by $e_\rho^u$ in the following. A $\rho$-measurement is a hidden measurement for the entity $S$.

Let us come to probability. The transition probabilities that the particle $P$ arrives at point $u$ (hence the outcome of the measurement is $o_1^u$) and $-u$ (hence the outcome of the measurement is $o_2^u$) under the influence of the measurement $e_\rho^u$, are respectively given by:

$$\mu_\rho(p_u, e_u, p_v) = \int_{-1}^{v \cdot u} \rho(x) dx$$
$$\mu_\rho(p_{-u}, e_u, p_v) = \int_{v \cdot u}^{1} \rho(x) dx$$
(4)

Let us now come to the Ellsberg paradox situation, and consider an urn with 90 balls of different colors, red, black and yellow. Let us assume that the pure state $p_v$ represents a physical situation where the number of balls is fixed, e.g., 30 red balls, 32 black balls and 28 yellow balls (another pure state is a physical situation where the number of balls is the same, i.e. 30). Thus, different combinations of colors correspond to different sectors on the sphere. Then, for each unit vector $u \in surf$, let us consider the measurement $e_u$ representing the decisional situation where the subject is asked to bet on a given color, red or black, and associate $e_u$ with the two outcomes $o_1^u$ and $o_2^u$ in such a way that, if the outcome $o_1^u$ ($o_2^u$) is obtained, this corresponds to the situation where the subject chooses to 'bet on red' ('bet on black'). Moreover, let us consider seven different kinds of elastics, characterized by the classical probability distributions, $\rho_1$, $\rho_2$, ..., $\rho_7$, one for each conceptual landscape defined in Sec. 2. The non-uniform distributions reflect the different cognitive aspects of the decisional process. Hence, the probabilities of 'betting on red' and 'betting on black' under the conceptual landscape represented by the classical probability distribution $\rho_j$ are respectively given by:

$$\mu_{\rho_j}(p_u, e_u, p_v) = \int_{-1}^{v \cdot u} \rho_j(x) dx$$
$$\mu_{\rho_j}(p_{-u}, e_u, p_v) = \int_{v \cdot u}^{1} \rho_j(x) dx$$
(5)

We note that, if we take into account the physical situation in which the urn contains 30 red balls, 30 black balls and 30 yellow balls, and locate the unit vector $v$ representing this physical situation in the north pole of the sphere, then, for every conceptual landscape, both probabilities in Eq. (5) are equal to 1/2, which corresponds to what one actually expects.

Till now we have considered only physical situations in which the number of balls was fixed, that is, the preparation of the balls in the urn was completely known. This situation was reflected by the fact that the physical state $p_v$ of the entity $S$ was a pure state and the point $v$ located on $surf$. But we know that in the Ellsberg paradox situation only the number of red balls is known, i.e. 30 balls, while black and yellow balls are in unknown proportion. This situation can be realized in our sphere model by introducing mixed states and representing them by inner points of the sphere. For example, if the subject knows that a physical situation associated with the pure state $p_v$ is mixed with a physical situation associated with the pure state $p_{-v}$ (where the point $-v \in surf$ is opposed to the point $v$) with probabilities $s \in [0,1]$ and $(1-s) \in [0,1]$, respectively, so that the state of the entity is a mixed state $p_w$, with $w = sv + (1-s)(-v)$, then the probabilities of 'betting on red' and 'betting on black' under the conceptual landscape represented by the classical probability distribution $\rho_j$ are respectively given by:

$$\mu_{\rho_j}(p_u, e_u, p_w) = p \int_{-1}^{v \cdot u} \rho_j(x) dx + (1-p) \int_{-1}^{-v \cdot u} \rho_j(x) dx$$
$$\mu_{\rho_j}(p_{-u}, e_u, p_w) = p \int_{v \cdot u}^{1} \rho_j(x) dx + (1-p) \int_{-v \cdot u}^{1} \rho_j(x) dx$$
(6)

The presentation of the first part of the Ellsberg paradox is thus completed.

It is important to observe that the probabilities in Eqs. (5) and (6) cannot be cast into a unique Kolmogorovian scheme, which can be proven by referring to *Pitowsky's polytopes*, or to *Bell-like inequalities* (Pitowsky 1989). Furthermore, if we limit ourselves to consider uniform probability distributions $\rho_j$, then the probabilities in Eqs. (5) and (6) become the standard quantum probabilities for spin measurements, since our sphere model is a model for a spin 1/2 quantum particle (Aerts 1993, 1995, 1998).

In the model illustrated above we limited ourselves to consider the first part of the Ellsberg paradox, namely the situation in which a subject is asked to decide between 'betting on red' and 'betting on black'. A more complex model should be constructed to take into account the whole paradox. We do not accomplish this task in the present paper, for the sake of lack of space. We instead observe that our simple sphere model already shows that the Ellsberg example cannot be modeled by using classical Kolmogorovian probabilities, because of its intrinsic contextuality, and that a non-Kolmogorovian quantum-like framework is necessary. This result is relevant in our opinion and we devote the next section to explain and clarify it.

## 4. Conclusions

The notion of contextual risk recently introduced by the authors (Aerts and Sozzo 2011) to mathematically represent ambiguity situations in economics has been successfully applied and particularized to the Ellsberg paradox. Within our hidden measurement formalism a sphere model for the Ellsberg situation has been elaborated which shows that a unique Kolmogorovian scheme is not suitable to model the experimental situation put forward by Ellsberg. Moreover, a quantum or quantum-like framework is needed because of the relevance of context in the form of conceptual landscapes in this situation. The analysis undertaken in this paper suggests the hypothesis that two structured and superposed layers can be identified in human thought: a *classical logical layer*, that can be modeled by using a classical Kolmogorovian probability framework, and a *quantum conceptual layer*, that can instead be modeled by using the probabilistic formalism of quantum mechanics. The thought process in the latter layer is given form under the influence of the totality of the surrounding conceptual landscape, hence context effects are fundamental in this layer (Aerts 2009).

Let us conclude this paper with two remarks. Firstly, we note that in our approach the explanation of the violation of the expected utility hypothesis and the Sure-Thing Principle is not (only) the presence of an ambiguity aversion. On the contrary, we argue that the above violation is due to the concurrence of superposed conceptual landscapes in human minds, of which some might be linked to ambiguity aversion, but other completely not. We therefore maintain that the violation of the Sure-Thing Principle should not be considered as a fallacy of human thought, as often claimed in the literature but, rather, as the proof that real subjects follow a different way of thinking than the one dictated by classical logic in some specific situations, which is context-dependent. Secondly, we observe that an explanation of the violation of the expected utility hypothesis and the Sure-Thing Principle in terms of quantum probability and quantum interference has already been presented in the literature (see, e.g., Busemeyer, Wang and Townsend 2006, Franco 2007, Khrennikov and Haven 2009, Pothos and Busemeyer 2009). What is new in our approach is the fact that the quantum mechanical modeling is not just an elegant formal tool but, rather, it reveals the presence of an underlying quantum conceptual thought. We stress, to conclude this paper, that the presence of a quantum structure in cognition and decision taking does not presuppose the presence of microscopic quantum processes in human mind. In fact, we have avoided making such a compelling assumption in our approach.